\documentclass[aps,prd,preprint,nofootinbib,superscriptaddress]{revtex4-1}
\usepackage{amsfonts}
\usepackage{mathrsfs}
\usepackage{graphicx}
\usepackage{amsmath}
\usepackage{amssymb}
\usepackage{multirow}
\usepackage{subfigure}
\usepackage{epsfig}
\usepackage{graphicx}
\usepackage{booktabs}
\usepackage{array}
\usepackage{tabularx}
\usepackage{slashed}

\newcommand{\bqa}{\begin{eqnarray}}
\newcommand{\eqa}{\end{eqnarray}}
\newcommand{\beq}{\begin{equation}}
\newcommand{\eeq}{\end{equation}}
\graphicspath{{fig/}{dia/}} \DeclareGraphicsExtensions{.eps}
\begin{document}
\baselineskip 20pt
\title{Production of doubly charmed tetraquark $T_{cc}$ via photon-photon fusion at electron-positron colliders}

\author{Jun Jiang}
\email{jiangjun87@sdu.edu.cn}
\affiliation{School of Physics, Shandong University, Jinan, Shandong 250100, China}

\author{Shi-Yuan Li}
\email{lishy@sdu.edu.cn}
\affiliation{School of Physics, Shandong University, Jinan, Shandong 250100, China}

\author{Xiao Liang}
\email{202120290@mail.sdu.edu.cn}
\affiliation{School of Physics, Shandong University, Jinan, Shandong 250100, China}

\author{Yan-Rui Liu}
\email{yrliu@sdu.edu.cn}
\affiliation{School of Physics, Shandong University, Jinan, Shandong 250100, China}

\author{Zong-Guo Si}
\email{zgsi@sdu.edu.cn}
\affiliation{School of Physics, Shandong University, Jinan, Shandong 250100, China}

\author{Zhong-Juan Yang}
\email{sps_yangzj@ujn.edu.cn}
\affiliation{School of Physics and Technology, University of Jinan, Jinan, Shandong 250022, China}

\begin{abstract}

Within a phenomenological diquark fragmentation model, we study the production of doubly charmed  tetraquark $T_{cc}$ via photon-photon fusion at electron-positron colliders.
The production of $T_{cc}$ is divided into two steps: the perturbative production of heavy $(cc)$-diquark and its nonperturbative hadronization.
Two diquark configurations of $(cc)[^3S_1]_{\bar{3}}$ and $(cc)[^1S_0]_{6}$ are considered, and the $(cc)[^3S_1]_{\bar{3}}$ state dominates the produciotn of $T_{cc}$.
We discuss two hadronization models of $(cc)[^3S_1]_{\bar{3}}$ intermediate state into the tetraquark $T_{cc}$.
It is found that it is promising to observe the tetraquark $T_{cc}$ via photon-photon fusion process both at the Circular Electron Positron Collider (CEPC) and the International Linear Collider (ILC).
We find that the cross sections are sensitive to constituent charm quark mass of diquark, 
and they also have strong dependence on the hadronization models.

\vspace {5mm} 
\end{abstract}

\maketitle

\section{INTRODUCTION}
\label{sec:introduction}

The past twenty years are the “golden age” for the heavy exotic hadrons which do not fit the conventional quark models. 
It begins with the discovery of $\chi_{c1}(3872)$ by Belle collaboration in the $J/\psi\pi^+\pi^-$ invariant mass spectrum in 2003 \cite{Belle:2003nnu}, which might be the candidate of high excited charmonium, compact tetraquark, charm-meson molecule or their mixture. 
Subsequently, the charged particle $Z_{c}^{+}(3900)$ is firstly observed by the BESIII collaboration in the $J/\psi\pi^{\pm}$ mass spectrum \cite{BESIII:2013ris}, which contains at least four quarks because it is hidden-charmed and charged. 
Recently, the doubly charmed tetraquark $T_{cc}^+(3875)$ (or concisely $T_{cc}$) with $cc\bar{u}\bar{d}$ constituent quarks is observed by the LHCb collaboration in the $D^0D^0\pi^+$ invariant mass spectrum \cite{LHCb:2021auc,LHCb:2021vvq}.
It has spin-parity quantum numbers $J^{P}=1^{+}$ and the data favors the isoscalar state. 
Then $T_{cc}$ has attracted significant attention of theoretical communities. 
For instance, the $T_{cc}$ state has been investigated using lattice QCD \cite{Lyu:2023xro,Padmanath:2022cvl,Chen:2022vpo}. 
Due to its proximity to the $D^{*+}D^0$ and $D^{*0}D^+$ thresholds, the internal structure of $T_{cc}$ can be interpreted as a charm-meson molecule 
\cite{Wang:2023ovj,Meng:2021jnw,Albaladejo:2021vln,Montesinos:2023qbx,Du:2021zzh,Braaten:2022elw}. 
However, $T_{cc}$ might also be the candidate of compact tetraquark with the $(cc)$-diquark configuration \cite{Jin:2021cxj,Qin:2020zlg,Weng:2021hje,Agaev:2021vur,Kim:2022mpa,Noh:2023zoq,Wu:2022gie,Li:2023wug,Dong:2024upa}. 
The nature of $T_{cc}$ is still under debat because of the missing of smoking gun.
In this manuscript, we consider $T_{cc}$ as a compact tetraquark.

We adopt a phenomenological diquark fragmentation model to describe the production of compact tetraquark $T_{cc}$. 
With the heavy $(cc)$-diquark constituent, the production of compact tetraquark $T_{cc}$ can be factored out into two steps: the perturbative production of the $(cc)$-diquark because of the non-relativistic nature of heavy diquark, and the nonperturbative hadronization of colored $(cc)$-diquark into $T_{cc}$.
In the $SU_C(3)$ color group, two charm quarks have the color decompositron $3 \otimes 3 = \bar{3} \oplus 6$, then the $(cc)$-diquark has the two spin-color configurations $[^3S_1]_{\bar{3}}$ and $[^1S_0]_6$ for the antisymmetry by exchanging two identical charm quarks.
In this manuscript, both two configurations are considered. 
This phenomenological diquark fragmentation model has been widely employed to study the doubly heavy baryons \cite{Ma:2003zk,Jin:2014nva,Li:2020ggh,Sun:2020mvl,Tian:2023uxe,Zhan:2023jfm,Yang:2024ysg} and $T_{cc}$ \cite{Yang:2024ysg,Niu:2024ghc}.

Both before and after the discovery of $T_{cc}$ at LHCb, there are  extensive phenomenological study to explore its nature.
We have pioneer works on the $QQ\bar{q}\bar{q}$ bound states \cite{Zouzou:1986qh,Manohar:1992nd}.
The double charm tetraquark states are analized under QCD sum rules \cite{Navarra:2012zz} and lattice QCD \cite{Collins:2024sfi}.
The production of $T_{cc}$ are investigated, $e.g.$ in $pp$ collision \cite{Chen:2011jtl,Ali:2018xfq,Ali:2018xfq,Qin:2020zlg,Hua:2023zpa}, at an electron-positron collider \cite{Hyodo:2012pm}, and in the decay of Higgs/$Z^0$/$W^+$ bosons \cite{Niu:2024ghc}.
In this work, we investigate the production of $T_{cc}$ via photon-photon fusion at an electron-positron collider.
The ongoing SuperKEKB collider with center-of-mass (CM) energy $\sqrt{s} = 10.6$ GeV and the future Circular Electron Positron Collider (CEPC) \cite{CEPCStudyGroup:2018rmc,CEPCStudyGroup:2018ghi} with $\sqrt{s} = 240$ GeV, and the International Linear Collider (ILC) \cite{ILC:2007oiw,ILC:2007bjz} with $\sqrt{s} = 500$ GeV are considered.

The remaining parts of this paper are organized as follows. 
In section \ref{sec:formulation}, we introduce the primary formulation and some technical details adopted in the diquark fragmentation model. 
In section \ref{sec:data}, we present the total cross sections and the differential distributions for the production of $T_{cc}$. Sec.\ref{sec:summary} is reserved for a summary.

\section{FORMULATION}
\label{sec:formulation}

\begin{figure}[!thbp]
    \centering    
    \includegraphics[width=0.8\textwidth]{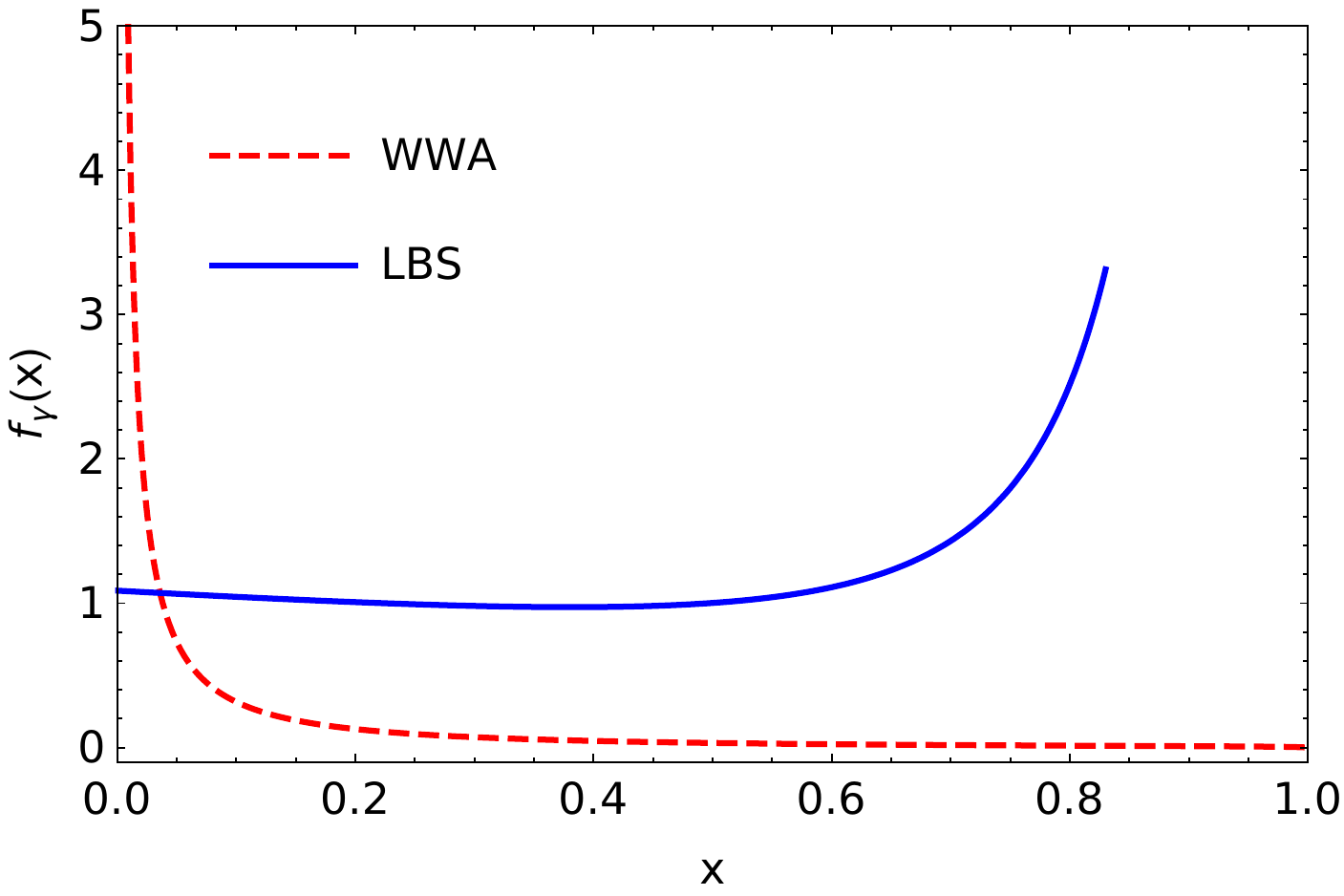}
    \caption{The WWA and LBS photon distribution functions in Eqs. \eqref{eq:wwa} and \eqref{eq:lbs}, respectively.}
    \label{fig:photonspectrum}
\end{figure}

The differential cross section for the production of doubly charmed tetraquark $T_{cc}$ via photon-photon fusion in electron-positron collision has the factored formula,
\begin{equation}
    d\sigma=\int dx_{1}dx_{2} f_{\gamma}(x_{1})f_{\gamma}(x_{2})d \hat{\sigma}( \gamma+\gamma\to T_{cc}[n]+\bar{c}+\bar{c}),
\end{equation}
where $f_{\gamma}(x)$ is the photon distribution function with $x=E_{\gamma}/E_{e}$ being the energy fraction of emitting photon from initial electrons or positrons, and $d \hat{\sigma}$ is the differential cross section for subprocess $\gamma+\gamma\to T_{cc}[n]+\bar{c}+\bar{c}$.
Initial photons can be generated through the bremsstrahlung effect which  is described by the Weizsacker-Williams approximation (WWA) \cite{vonWeizsacker:1934nji,Williams:1934ad,Frixione:1993yw},
\begin{equation}
    f^{\mathrm{WWA}}_{\gamma}(x) = \frac{\alpha}{2\pi}\left(\frac{1+(1-x)^2}{x}\log\left(\frac{Q^{2}_{\rm max}}{Q^{2}_{\rm min}}\right)+2m^2_{e}x\left(\frac{1}{Q^{2}_{\rm max}}-\frac{1}{Q^{2}_{\rm min}}\right)\right),
    \label{eq:wwa}
\end{equation}
where $Q^{2}_{\rm min}=m^{2}_{e}x^{2}/(1-x)$, $Q^{2}_{\rm max}=(\theta_{c}\sqrt{s}/2)^2(1-x)+Q^{2}_{\rm min}$, $m_e$ is the mass of electron, $s$ is the squared CM energy, and $\theta_{c}=32$ mrad is the maximum scattering angle of the electron or positron \cite{Klasen:2001cu}. 
We adopt the WWA spectrum for the photon at SuperKEKB with CM energy $\sqrt{s} = 10.6$ GeV and CEPC with CM energy $\sqrt{s} = 240$ GeV.
Another source of photon-photon scattering is the laser back scattering (LBS) effect. It was proposed for the linear $e^+e^-$ accelerators like ILC with CM energy $\sqrt{s} = 500$ GeV and higher. The LBS photon energy spectrum is \cite{Ginzburg:1981vm}
\begin{equation}
    f^{\mathrm{LBS}}_{\gamma}(x)=\frac{1}{N}\left(1-x+\frac{1}{1-x}-\frac{4x}{x_{m}(1-x)}+\frac{4x^2}{x_{m}^2(1-x)^2}\right),
    \label{eq:lbs}
\end{equation}
where $x_{m} \approx 4.83$ \cite{Telnov:1989sd}, the energy fraction $x$ is restricted by $0 \leq x \leq \frac{x_m}{1+x_m}\approx 0.83$, and the normalization factor $N$ is
\begin{equation}
    N=\left(1-\frac{4}{x_{m}}-\frac{8}{x^{2}_{m}}\right)\log(1+x_{m})+\frac{1}{2}+\frac{8}{x_{m}}-\frac{1}{2(1+x_{m})^2} .
\end{equation}
We compare the two photon distribution functions of Eq. \eqref{eq:wwa} and Eq. \eqref{eq:lbs} in Fig. \ref{fig:photonspectrum}.

\begin{figure}[!thbp]
    \centering    
    \includegraphics[width=0.9\textwidth]{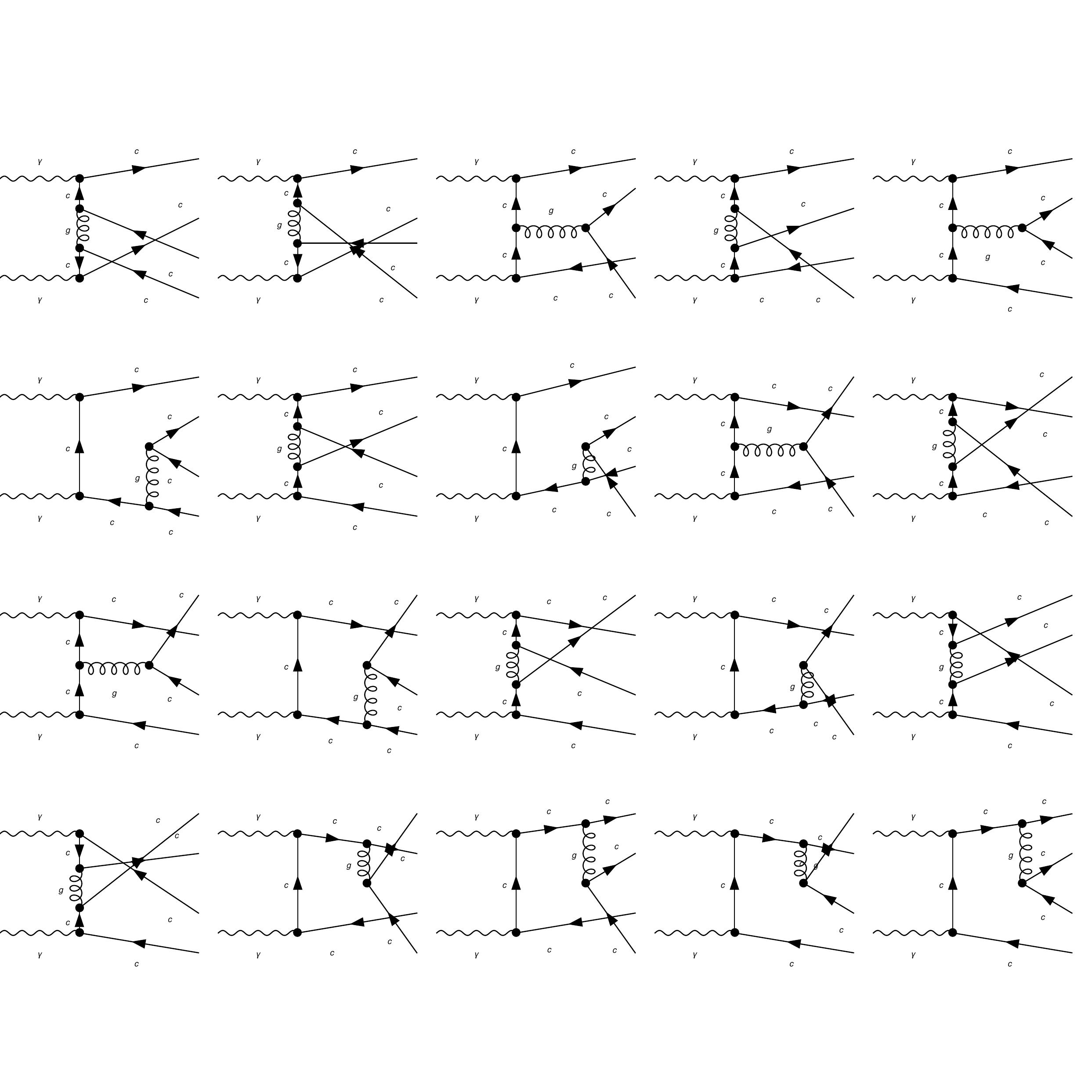}
    \caption{Typical twenty Feynman diagrams for $\gamma+\gamma\to (cc)[n]+\bar{c}+\bar{c}$ subprocess. Among the final states, the above two charm quarks will be bounded into the diquark $(cc)[n]$. The rest twenty diagrams can be obtained by interchanging two initial photon lines.}
    \label{feynman}
\end{figure}

Within the framework of diquark fragmentation model, the cross section for subprocess $\gamma+\gamma\to T_{cc}[n]+\bar{c}+\bar{c}$ can be factorized into two parts: the short-distance coefficient for the production of a diquark $(cc)[n]$, and the long-distance transition probability from the diquark state $(cc)[n]$ hadronization to the tetraquark $T_{cc}[n]$,
\begin{equation}
d\hat{\sigma}( \gamma+\gamma\to T_{cc}[n]+\bar{c}+\bar{c})  \\
= d\hat\sigma(\gamma+\gamma\to (cc)[n]+\bar{c}+\bar{c}) \times {\cal P}\left((cc)[n] \to T_{cc}[n] \right),     \label{dsigma}
\end{equation}
where the short-distance coefficient $d\hat\sigma(\gamma+\gamma\to (cc)[n]+\bar{c}+\bar{c})$ can be calculated perturbatively, the long-distance transition probability ${\cal P}\left((cc)[n] \to T_{cc}[n] \right)$ represents the hadronization probability of the $(cc)[n]$ diquark into the $T_{cc}[n]$ tetraquark state rather than baryons like $\Xi_{cc},\,\Omega_{cc}$ or others, and $[n]$ stands for the spin and color states $[^1S_0]_{6}$ or $[^3S_1]_{\bar{3}}$.
The contribution of Eq. \eqref{dsigma} can be classified as the usual fragmentation mechanism, specifically the diquark fragmentation here, so we call it the diquark fragmentation model in this paper. 
This contribution becomes dominant only in the large transverse momentum $p_t$ region, which will be further discussed in numerical results. 
In contrast, there are contributions from other production mechanisms, for example, all the quark constitutes $cc\bar{u}\bar{d}$ of $T_{cc}$ are produced at short-distance first then followed by the hadronization, which would dominate in the relatively small $p_t$ region. 
In this work, these production mechanisms are neglected.

The short-distance differential cross section for the production of diquark $d\hat\sigma(\gamma+\gamma\to (cc)[n]+\bar{c}+\bar{c})$ has the general formula,
\begin{eqnarray}
d\hat\sigma(\gamma+\gamma\to (cc)[n]+\bar{c}+\bar{c})=\frac{1}{4\sqrt{(p_1\cdot p_2)^2-m^4_{e}}} \overline{\sum}  \big|{\cal M}\big|^{2} d\Phi_3,
\label{sd-sigma}
\end{eqnarray}
where $\cal M$ is the amplitude for the production of $(cc)[n]$ diquark, $\overline{\sum}$ stands for the average over the spins of the initial states and the sum over colors and spins of the final states. 
There are forty Feynman diagrams for the $\gamma+\gamma\to (cc)[n]+\bar{c}+\bar{c}$ subprocess at leading order in ${\cal O}(\alpha_s^2)$. Half of them are presented in Fig. \ref{feynman}, and the rest can be obtained by interchanging two initial photon lines.
Note that, an extra factor of $\frac{1}{2!2!}$ shall be multiplied in the squared amplitudes due to the symmetry of identical particles.
In the photon-photon CM frame, the three-body phase space $d\Phi_3$ can be simplified as
\begin{eqnarray}
    d{\Phi_3} &=& (2\pi)^4 \delta^{4}\left(p_1+p_2 - \sum_{f=1}^3 q_{f}\right)\prod_{f=1}^3 \frac{d^3{\vec{q}_f}}{(2\pi)^3 2q_f^0} \nonumber \\
    &=& d s_2 d\Phi_2(s,m_1^2,s_2) d\Phi_2(s_2,m_2^2,m_3^2),
    \label{dPhi-3}
\end{eqnarray}
where $s_2 = (q_2+q_3)^2$, the differential two-body phase space $d\Phi_2(s,m_1^2,s_2)= \frac{\sqrt{\lambda(s,m_1^2,s_2)}}{8s} d\Omega$ with $\lambda(x,y,z) = (x-y-z)^2-4yz$, the $\Omega$ is the solid angle, and $m_i \, (i=1,2,3)$ are the masses of final states.

\subsection{Production of $(cc)[n]$ diquark}
The scattering amplitude $\cal M$ for the production of heavy diquark can be obtained by a proper transformation from the scattering amplitude for the production of heavy quarkonium \cite{Jiang:2012jt}. 
The scattering amplitude for the production of $(c\bar{c})[n]$ charmonium in $\gamma(p_1)+\gamma(p_2)\to (c\bar{c})[n](q_1)+\bar{c}(q_2)+c(q_3)$ subprocess has the following form under the non-relativistic QCD (NRQCD) framework \cite{Bodwin:1994jh,Petrelli:1997ge},
\begin{eqnarray}
    {\cal M}\left((c\bar{c})[n]\right) &=& \varepsilon(p_1)\varepsilon(p_2) \times \bar{u}(q_{11}) \cdots v(q_{2})\times \bar{u}(q_{3}) \cdots v(q_{12}) \times {\cal B}(q_1,M_{c\bar{c}}) \times {\cal C},
\label{mesonline}
\end{eqnarray}
where Lorentz indices are omitted, $\varepsilon(p_1)$ and $\varepsilon(p_2)$ are the polarization vectors of the initial photon, $u$ and $v$ are spinors, dots “$\cdots$” represent sequential interaction vertexes and fermion or gluon propagators, ${\cal B}(q_1,M_{c\bar{c}})$ is the wave function which contains the information of hadronization, $\cal C$ is the color factor of $(c\bar{c})[n]$ meson, and $M_{c\bar{c}}$ is the $(c\bar{c})[n]$ meson mass.
In contrast, the scattering amplitude for the production of $(cc)[n]$ diquark in $\gamma(p_1)+\gamma(p_2)\to (cc)[n](q_1)+\bar{c}(q_2)+\bar{c}(q_3)$ subprocess can be written as
\begin{eqnarray}
    {\cal M}\left((cc)[n]\right) &=& \varepsilon(p_1)\varepsilon(p_2) \times \bar{u}(q_{11}) \cdots v(q_{2})\times \bar{u}(q_{12}) \cdots v(q_{3}) \times {\cal B}^{\prime}(q_1,M_{cc}) \times {\cal C^{\prime}},
\label{diquarkline}
\end{eqnarray}
where ${\cal B}^{\prime}(q_1,M_{cc})$ and $\cal C^{\prime}$ are the wave function and color factor of $(cc)[n]$ diquark respectively, and $M_{cc}$ is the $(cc)[n]$ diquark mass. 

By comparing Eq. \eqref{mesonline} and Eq. \eqref{diquarkline}, we see that the two scattering amplitudes have different wave functions and color factors, and one of the Dirac fermion chains is reversed.
The wave functions and color factors are overall factors.
Then one can reverse the Dirac fermion chain to correlate the scattering amplitudes of diquark and charmonium.
The following formula for charge conjugation $C$ are helpful,
\begin{eqnarray}
&& v^T C = -\bar{u},\;\;\;C^- \Gamma^T_i C = -\Gamma_i,\;\;\;CC^- = 1, \nonumber \\
&& C^-s^T_f(k,m)C = s^T_f(-k,m),\;\;\; C^-\bar{u}^T = v,
\end{eqnarray}
where $\Gamma_i$ is the interaction vertex, $s_f(k,m)$ is the propagator with momentum $k$ and mass $m$, and superscripts $T$ and $-$ stand for the transpose and inverse operations, respectively.
Then the scattering amplitude for the production of $(cc)[n]$ diquark can be transformed into
\begin{eqnarray}
    {\cal M}\left((cc)[n]\right) &=& (-1)^{(\rho+1)}  \varepsilon(p_1)\varepsilon(p_2) \times \bar{u}(q_{11}) \cdots v(q_{2})\times \bar{u}(q_{3}) \cdots v(q_{12})\nonumber \\
    && \times {\cal B}^{\prime}(q_1,M_{cc}) \times {\cal C^{\prime}},
    \label{diquarkline2}
\end{eqnarray}
which has the same Dirac fermion chains as the one for charmonium in Eq. \eqref{mesonline}, and $\rho$ is the number of interaction vertex $\Gamma_i$ in the reversed $\bar{u}(q_{3}) \cdots v(q_{12})$ fermion chain. 
Thus, one can manipulate the scattering amplitudes of diquark in a similar way as those of charmonium.

Being analogous to the charmonium, the momenta of two constituent charm quarks of the $(cc)[n]$ diquark $q_{11}$ and $q_{12}$ have the following simple form,
\begin{equation}
    q_{11} = \frac{1}{2} q_1+q \;\;\;\; q_{12} = \frac{1}{2} q_1-q,
\end{equation}
where $q_1$ is the momentum of the $(cc)[n]$ diquark, and $q$ is the relative momentum between the two constituent charm quarks. 
The relative momentum $q$ is set to be zero at the leading velocity expanding in NRQCD.

The projector for two constituent charm quarks into $(cc)$-diquark with definite spin and color quantum numbers take the following replacement \cite{Petrelli:1997ge},
\begin{equation}
\label{projector}
v(q_{12})\bar{u}(q_{11}) \longrightarrow \frac{1}{2 \sqrt{2m_c}} \left(\xi_1\gamma^{5}+\xi_2\slashed{\epsilon}(q_1)\right)(\slashed{q}_1+2m_c)\otimes {\cal G} ,
\end{equation}
where $\xi_1=1$ and $\xi_2=0$ correspond to spin-singlet $[^1S_0]$ state of $(cc)[n]$ diquark, and $\xi_1=0$ and $\xi_2=1$ represent the spin-triplet $[^3S_1]$ case. $\epsilon(q_1)$ is the polarization vector of spin-triplet state, and ${\cal G}$ is the color factor of the diquark. Note, the mass of diquark $2m_c$ can be different from the mass $m_{T_{cc}}$ of $T_{cc}$ because the subsequent hadronization of diquark can further generat the mass.

The $(cc)$-diquark has two independent color states, anti-triplet ${ \bar{3}}$ color state and sextuplet $6$ color state, which are derived from the $SU_C(3)$ color charge group $3 \bigotimes 3={ \bar{3}} \bigoplus { 6}$. 
The color factor $\cal C^\prime$ in Eq. \eqref{diquarkline2} has the following form
\begin{eqnarray}
{\cal C^\prime}_{ijk}={N}\times\sum_{m,n}(T^{a})_{im}(T^{a})_{jn} \times {\cal G}_{mnk},
\end{eqnarray}
where $i,j,m,n$ are the color indices of the two outgoing anti-charm quarks and two constituent charm quarks of the diquark respectively, and $k$ is the color index of the diquark $(cc)[n]$, 
$T^a$ are the color matrix from the gluons with $a=1,\cdots,8$, 
${{N}}=\sqrt{1/2}$ is the normalization constant, and ${\cal G}_{mnk}$ is the color factor for diquark in Eq. \eqref{projector}. 
The function ${\cal G}_{mnk}$ adopts the anti-symmetric function $\varepsilon_{mnk}$ when diquark is anti-triplet ${ \bar{3}}$ color state, while it adopts the symmetric function $f_{mnk}$ when diquark is sextuplet ${ 6}$ color state. 
The anti-symmetric function $\varepsilon_{mnk}$ and the symmetric function $f_{mnk}$ satisfies following rules,
\begin{eqnarray}
    \varepsilon_{mnk}\varepsilon_{m'n'k} &=\delta_{mm'}\delta_{nn'}-\delta_{mn'}\delta_{nm'} ,\\
    f_{mnk}f_{m'n'k}&=\delta_{mm'}\delta_{nn'}+
\delta_{mn'}\delta_{nm'}.
\end{eqnarray}
Then we have the squared and summed color facor ${\cal C}^{\prime 2}_{ijk} = \frac{4}{3}$ for anti-triplet ${\bar{3}}$ color state, and ${\cal C}^{\prime 2}_{ijk} = \frac{2}{3}$ for sextuplet ${6}$ color state.

\subsection{Hadronization from $(cc)[n]$ diquark to tetraquark $T_{cc}$}
\label{subsec:transition}
The the long-distance hadronization probability ${\cal P}\left((cc)[n] \to T_{cc}[n] \right)$ in Eq. \eqref{dsigma} describes the hadronization from intermediate $(cc)[n]$ diquark in spin and color state $[n]$ to the color singlet tetraquark $T_{cc}$. Usually, the probability ${\cal P}\left((cc)[n] \to T_{cc}[n] \right)$ can be determined by fitting the experimental results, or the potential models.
In this paper, we adopt two different hadronization schemes to estimate ${\cal P}\left((cc)[n] \to T_{cc}[n] \right)$.

By adopting the potential models, the hadronization probability can be related to the Sch$\ddot{\mathrm{o}}$dinger wave function at the origin $\Psi_{T_{cc[n]}}(r=0)$, which can further be related to the radial wave function at origin for S-wave \cite{Hyodo:2012pm},
\begin{equation}
{\cal P}\left((cc)[n] \to T_{cc}[n] \right) \longrightarrow |\Psi_{T_{cc[n]}}(r=0)|^2 = \frac{|R_{T_{cc[n]}}(r=0)|^2}{4\pi},
\end{equation}
where $R_{T_{cc[n]}}(r)$ is the radial wave function with the normalization $\int_0^{\infty} r^2 d r |R_{T_{cc[n]}}(r)|^2=1$. 
And the radial wave functions at origin can be obtained by solving Sch$\ddot{\mathrm{o}}$dinger equation under the specified potentials. 
The method and the Mathematica code are described in Ref. \cite{Lucha:1998xc}. 
In this manuscript, we adopt the harmonic oscillator potential (HOP) \cite{Hyodo:2012pm},
\begin{equation}
V= {\textstyle \sum_{i<j}}(-\frac{3}{16})\lambda_i \cdot \lambda_j \frac{k}{2} \left | r_i r_j \right | ^2,
\end{equation}
where $r_i$ is the position of quark $i$, $\lambda_i \cdot \lambda_j$ is the color factor which has different forms for each color channel $\bar{3}$ and $6$, and $k$ is the strength parameter of the harmonic oscillator potential for quark confinement. $k=0.33 \,\mathrm{GeV}^3$ is fixed to reproduce the value of the wave function for charmonia under the Cornel potential model. More details can be found in Ref. \cite{Hyodo:2012pm}, and we directly adopt their estimates for the two configurations, 
\begin{eqnarray}
    {\cal P}\left((cc)[^1S_0]_{6} \to T_{cc}[^1S_0]_{6} \right) &\longrightarrow 0.054\;{\rm GeV}^3, \\
    {\cal P}\left((cc)[^3S_1]_{\bar{3}} \to T_{cc}[^3S_1]_{\bar{3}} \right) &\longrightarrow 0.089 \;{\rm GeV}^3.
\end{eqnarray}

The heavy diquark-antiquark symmetry (HDAS) \cite{Lichtenberg:1989ix,Anselmino:1992vg,Carlson:1987hh,Savage:1990di,Brambilla:2005yk,Fleming:2005pd,Cohen:2006jg} indicates that heavy diquark ${(QQ)}$ with color state $\bar{3}$ can be related with antiquark ${\bar{Q}}$ with the same color. 
So the hadronization of diquark $(cc)_{\bar{3}}$ with anti-triplet color into tetraquarks state $T_{cc}$ with quark constituents $cc\bar{u}\bar{d}$ can be related to the transition of an anti-charm quark $\bar{c}_{\bar{3}}$ with anti-triplet color to heavy antibaryon 
$\bar{\Lambda}_c^-$ with quark constituents $\bar{c}\bar{u}\bar{d}$,
\begin{equation}
(cc)_{\bar{3}}\to T_{cc}(cc\bar{u}\bar{d}) \Longleftrightarrow \bar{c}_{\bar{3}} \to \bar{\Lambda}_c^-(\bar{c}\bar{u}\bar{d}).
\end{equation}
And the transition probability of $\bar{c} \to \bar{\Lambda}_c^-(\bar{c}\bar{u}\bar{d})$ process can be estimated by the fragmentation fraction of $c \to \Lambda_c^+ (cud)$, whose value can be obtained from the fitting to experimental data $f(c \to \Lambda_c^+)=0.0623$ \cite{Zenaiev:2016qcf}.
In this HDAS scheme, it takes two steps from the heavy $cc$ pair to tetraquark $T_{cc}$ \cite{Niu:2024ghc,Chen:2011jtl}. 
First, we have the Sch$\ddot{\mathrm{o}}$dinger wave function at the origin $\Psi_{cc}(0)$ which describes the probability that $cc$ pair is bounded into the S-wave diquark, whose value can be predicted by solving Sch$\ddot{\mathrm{o}}$dinger equation. 
Under the Coulomb potential, $|\Psi_{cc}(0)|^2 = 0.0198 \;{\rm GeV}^3$ is estimated in Ref. \cite{Chen:2011jtl}. 
Under the pow-low potential \cite{Bagan:1994dy}, $|\Psi_{cc}(0)|^2 = 0.039 \;{\rm GeV}^3$ is estimated in Ref. \cite{Baranov:1995rc}, which is adopted in our calculation.
Second, the diquark hadronizes into $T_{cc}$ by the fragmentation function of $f((cc)_{\bar{3}}\to T_{cc}) \approx f(c \to \Lambda_c^+)$.
So the hadronization probability for $[^3S_1]_{\bar{3}}$ configuration takes the following replacement,
\begin{equation}
     {\cal P}\left((cc)[^3S_1]_{\bar{3}} \to T_{cc}[^3S_1]_{\bar{3}} \right) \longrightarrow |\Psi_{cc}(0)|^2  \times f(c \to \Lambda_c^+) = 0.00243 \;{\rm GeV}^3.
\end{equation}
We remind the readers that this transition probability in HDAS scheme is about 2.7\% of that in HOP scheme, which will result in large suppression on the cross sections.
This indicates the large dependence of our results on the hadronization schemes.
Anyway, the non-perturbative hadronization is one of the key problem to describe the formation of exotic hadrons in QCD.
Very recently, the Born-Oppenheimer potentials for QCD are used to obtain the $T_{QQ}$ wave function with no need to assume a model \cite{Berwein:2024ztx,Braaten:2024tbm,Brambilla:2024thx}.

\section{RESULTS}
\label{sec:data}

In the numerical evaluation, other input parameters are listed below,
\begin{eqnarray}
& & \alpha=1/137.065,\; \alpha_{s}(\mu)=\frac{4\pi}{\beta_0\ln{\mu^2/\Lambda^2_{\rm QCD}}},\; m_{T_{cc}}=3.88 \; {\rm GeV}.
\end{eqnarray}
Here, $\beta_0=\tfrac{11}{3}C_A-\tfrac{4}{3}T_Fn_f$ with $n_f=4$, $\Lambda_{\rm QCD}=297\ {\rm MeV}$, and $\mu=\sqrt{4m_c^2+p_t^2}$ with $p_t$ being the transver momentum of the tetraquark.
For the non-perturbative transition probability ${\cal P}\left((cc)[n] \to T_{cc}[n] \right)$, we adopt the values in the two different hadronization schemes as mentioned in Sec. \ref{subsec:transition}. 
The harmonic oscillator potential (HOP) for both $T_{cc}[^1S_0]_6$ and $T_{cc}[^3S_1]_{\bar{3}}$ configurations, and the fragmentation method under the heavy diquark-antiquark symmetry (HDAS) for $T_{cc}[^3S_1]_{\bar{3}}$ only.

\begin{table}[!thbp]
    \caption{The cross sections in unit of $10^{-3}$ fb for $T_{cc}[^1S_0]_{6}$ and $T_{cc}[^3S_1]_{\bar{3}}$ production in two hadronization schemes with WWA photon spectrum at the SuperKEKB. Here $\sqrt{s}=10.6\ {\rm GeV}$, the transverse momentum cut $ 0.2\ {\rm GeV} \le p_{t} \le 4.0\ {\rm GeV}$ is employed to $T_{cc}$. $m_{T_{cc}}=3.88$ GeV is fixed, but with variable masses of constituent charm quark of $(cc)[n]$ diquark $m_c=1.94$, 1.7 and 1.5 GeV.}
    \begin{center}       
       \begin{tabular}{p{2cm}<{\centering} p{4cm}<{\centering} p{4cm}<{\centering} p{4cm}< {\centering} } 
        \botrule
             SuperKEKB  & $T_{cc}[^1S_0]_{6}$   &  $T_{cc}[^3S_1]_{\bar{3}}$ & $T_{cc}[^3S_1]_{\bar{3}}$ \\
            &(HOP)&(HOP)&(HDAS)\\
        \hline
        $m_c = 1.94$ &
        $0.13$ &
        $20.94$ &
        $0.57$ \\ \hline
        $m_c = 1.7$ &
        $0.46$ &
        $58.44$ &
        $1.60$ \\  \hline
        $m_c = 1.5$ &
        $1.24$ &
        $134.94$ &
        $3.68$ \\
        \botrule
      \end{tabular}
    \end{center}
    \label{KEKB}
\end{table}

\begin{table}[!thbp]
    \caption{The cross sections in unit of fb for $T_{cc}[^1S_0]_{6}$ and $T_{cc}[^3S_1]_{\bar{3}}$ production in two hadronization schemes. For the CEPC with $\sqrt{s}=240\ {\rm GeV}$ using WWA spectrum, the transverse momentum cut $ 1.0\ {\rm GeV} \le p_{t} \le 50.0\ {\rm GeV}$ is employed to $T_{cc}$. For the ILC with $\sqrt{s}=500\ {\rm GeV}$ using LBS spectrum, the transverse momentum cut $ 1.0\ {\rm GeV} \le p_{t} \le 100.0\ {\rm GeV}$ is employed to $T_{cc}$. $m_{T_{cc}}=3.88$ GeV is always fixed, but with variable masses of constituent charm quark of $(cc)[n]$ diquark $m_c=1.94$, 1.7 and 1.5 GeV.}
    \begin{center}       
       \begin{tabular}{p{2cm}<{\centering} p{4cm}<{\centering} p{4cm}<{\centering} p{4cm}< {\centering} } 
        \botrule
            CEPC  & $T_{cc}[^1S_0]_{6}$   &  $T_{cc}[^3S_1]_{\bar{3}}$ & $T_{cc}[^3S_1]_{\bar{3}}$ \\
            &(HOP)&(HOP)&(HDAS)\\
        \hline
        $m_c=1.94$ &
        $3.46$ &
        $124.94$ &
        $3.41$ \\
        \hline
        $m_c=1.7$ &
        $5.42$ &
        $181.08$ &
        $4.94$ \\
        \hline
        $m_c=1.5$ &
        $8.47$ &
        $261.52$ &
        $7.14$ \\
        \botrule
        ILC  & $T_{cc}[^1S_0]_{6}$   &  $T_{cc}[^3S_1]_{\bar{3}}$ & $T_{cc}[^3S_1]_{\bar{3}}$ \\
        &(HOP)&(HOP)&(HDAS)\\
        \hline
        $m_c=1.94$ &
        $24.85$ &
        $416.39$ &
        $11.37$ \\
        \hline
        $m_c=1.7$ &
        $35.15$ &
        $577.87$ &
        $15.78$ \\
        \hline
        $m_c=1.5$ &
        $49.88$ &
        $801.17$ &
        $21.87$ \\
        \botrule
      \end{tabular}
    \end{center}
    \label{CEPC}
\end{table}

We first discuss the total cross section for the production of tetraquark $T_{cc}$ via photon-photon fusion in $e^+e^-$ collision. 
At the SuperKEKB collider whose CM collision energy is $10.6\ {\rm GeV}$, the cross sections are estimated using the WWA photon spectrum which are presented in Table \ref{KEKB}. 
While in Table \ref{CEPC}, the WWA photon spectrum is used at the CEPC with CM energy $\sqrt{s} = 240$ GeV, and the LBS photon spectrum is used at ILC with CM energy $\sqrt{s} = 500$ GeV. 
It is worthy noting that different transverse momentum cuts are adopted in the evaluation of total cross sections at different colliders.
We find that, under the same hadronization scheme, the contribution from $[^3S_1]_{\bar{3}}$ state dominates the production of $T_{cc}$, while contribution from $[^1S_0]_6$ is negligible.
And the contributions from $[^3S_1]_{\bar{3}}$ state in HOP hadronization scheme are much greater than those in HDAS hadronization scheme because of the much greater hadronization probability.
In these two tables, the mass of $T_{cc}$ is always fixed at $m_{T_{cc}}=3.88$ GeV, but the mass of constituent charm quark of $(cc)[n]$ diquark takes three options, $m_c=1.94$, 1.7 and 1.5 GeV.
Since the hadronization of diquark into tetraquark can provide extra mass to $m_{T_{cc}}$, it is reasonable that $m_c$ can takes values smaller than half of the $T_{cc}$ mass.
It is also found that the drop in charm quark mass will lead to significant increases in cross sections for both $T_{cc}[^1S_0]_{6}$ and $T_{cc}[^3S_1]_{\bar{3}}$ states.
This phenomenon is more obvious at SuperKEKB than at CEPC and at ILC because the CM energy at SuperKEKB is more close to the threhold $m_{T_{cc}}+2m_c$.

Considering the uncertainties from both the mass variation and hadronization schemes, at the SuperKEKB collider with the luminosity of final design parameter of $8 \times 10^{35}\ {\rm cm}^{-2}{\rm s}^{-1}$ \cite{Zhou:2023dhi}, we have about $5 \sim 10^4$ events of $T_{cc}$ per year.
Taking a typical luminosity value of $10^{34}\ {\rm cm}^{-2}{\rm s}^{-1}$ at the CEPC \cite{CEPCStudyGroup:2023quu}, the $T_{cc}$ events are expected to be $(0.7 \sim 27.0)\times 10^{3}$ per year. 
Taking a typical luminosity value of $2 \times 10^{34}\ {\rm cm}^{-2}{\rm s}^{-1}$ at the ILC \cite{ILC:2007oiw}, the $T_{cc}$ events are expected to be $(0.7 \sim 17.0)\times 10^{4}$ per year. 
Considering the reconstruction of $T_{cc}$ by $D^0D^0\pi^+$ with approximated 100\% and the branching fraction $Br(D^0 \longrightarrow K^- \pi^+) = 3.95\%$ \cite{ParticleDataGroup:2024cfk}.
These results show that the experimental study of $T_{cc}$ at CEPC and ILC is promising.

\begin{figure}[!thbp]
    \centering
    \includegraphics[width=0.8\textwidth]{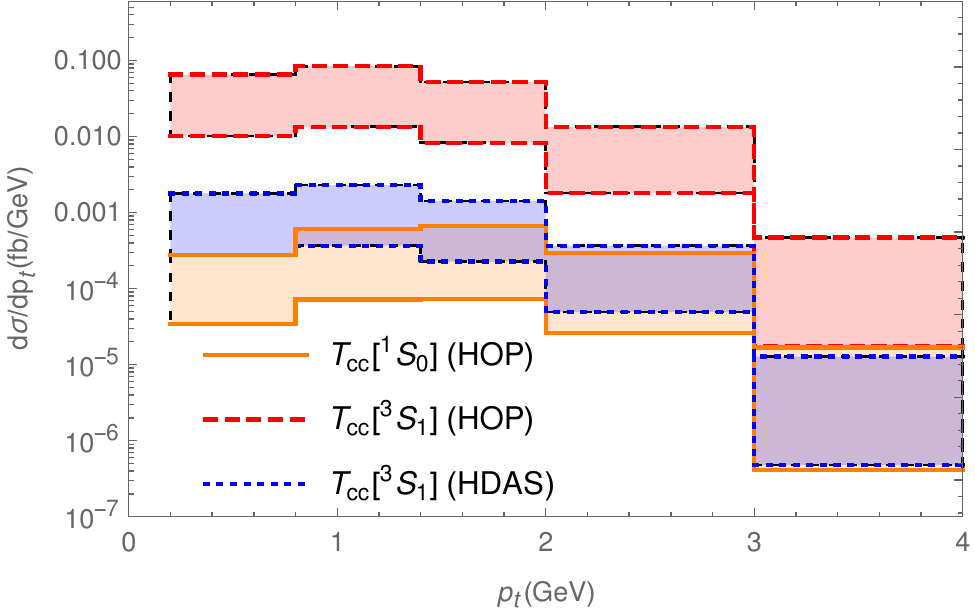}
    \caption{The $p_t$ distribution for the $T_{cc}$ production in two hadronization schemes with WWA photon spectrum at the SuperKEKB. Here $\sqrt{s}=10.6\; {\rm GeV}$, $m_{T_{cc}}=3.88$ GeV and the band is caused by the $m_c \in [1.5, 1.94]\; {\rm GeV}$.}
    \label{KEKBpt}
\end{figure}

\begin{figure}[!thbp]
    \centering
    \includegraphics[width=0.49\textwidth]{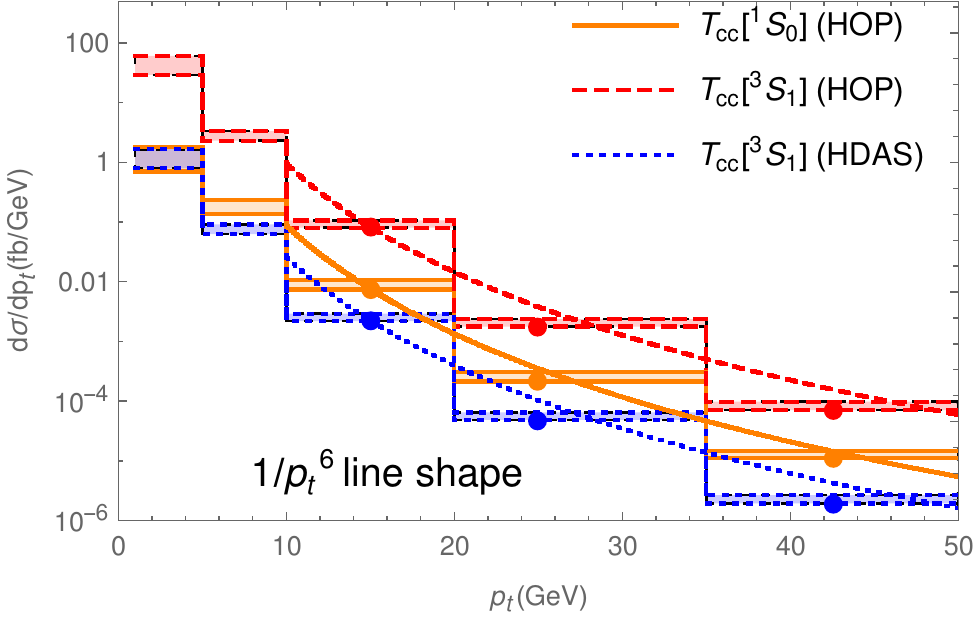}
    \includegraphics[width=0.49\textwidth]{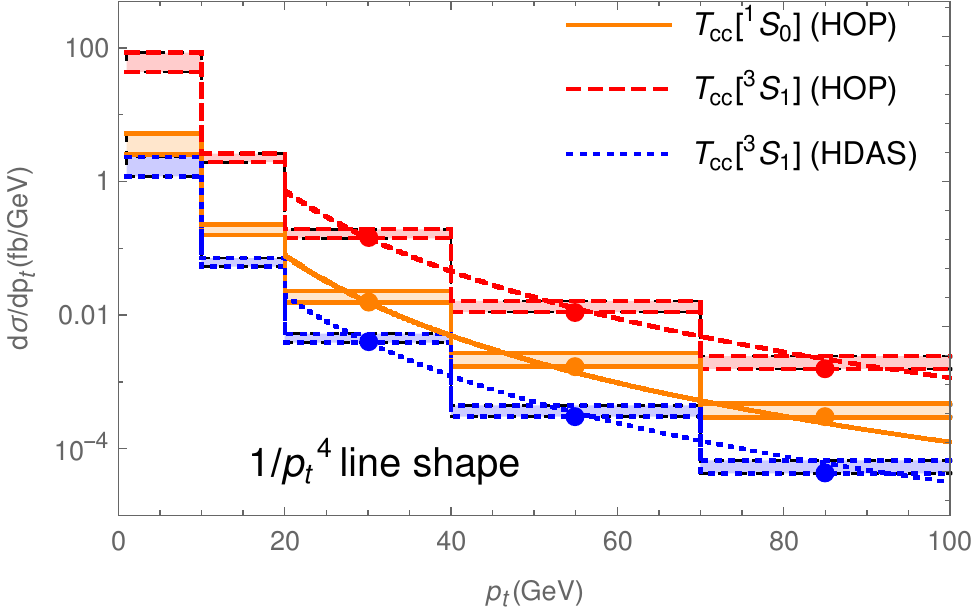}
    \caption{The $p_t$ distribution for the $T_{cc}$ production in two hadronization models at the CEPC (left panel) and ILC (right panel). The last three data points of each curve are used to fit the $1/p_t^6$ and $1/p_t^4$ line shapes for CEPC and ILC, respectively. Here $\sqrt{s}=240\; {\rm GeV}$ for CEPC and $\sqrt{s}=500\; {\rm GeV}$ for ILC, $m_{T_{cc}}=3.88$ GeV and the band is caused by the $m_c \in [1.5, 1.94]\; {\rm GeV}$.}
    \label{CEPCpt}
\end{figure}

We then study the transverse momentum $p_t$ distribution of $T_{cc}$.
In Fig. \ref{KEKBpt} and Fig. \ref{CEPCpt}, we display the transverse momentum $p_t$ distributions of $T_{cc}$ at SuperKEKB and at CEPC/ILC, respectively. 
In these figures, the mass $m_{T_{cc}}=3.88$ GeV of $T_{cc}$ is always fixed, and the uncertainty band is caused by the variance of the constituent charm quark in the diquark $m_c \in [1.5, 1.94]\; {\rm GeV}$. 
At SuperKEKB, each line shape exhibits a mild bulge within the bins of $0.2 \sim 2\ {\rm GeV}$ and then decreases. 
While at the CEPC (left panel in Fig. \ref{CEPCpt}) and ILC (right panel in Fig. \ref{CEPCpt}), the differential cross sections decrease monotonously. 
In the inclusive quarkonium production, the $p_t$ scaling of the partonic cross section at large $p_t$ region scales with $1/p_t^4$ for the single parton fragmentation, while it scales with $1/p_t^6$ for the double parton fragmentation \cite{Lansberg:2019adr,Kang:2014tta}.
We use the data of CEPC and ILC at large $p_t$ regions to fit both $1/p_t^4$ and $1/p_t^6$ line shapes. 
It is found that the CEPC data at $10 \,\mathrm{GeV}<p_t<50 \,\mathrm{GeV}$ scales with the $1/p_t^6$ line shape, while the ILC data at $20 \,\mathrm{GeV}<p_t<100 \,\mathrm{GeV}$ scales with the $1/p_t^4$ line shape. 
This is understandable. First, the $p_t$ of $T_{cc}$ at ILC is larger than that at CEPC, because the CM energy at ILC is about two times of that at CEPC and the LBS photon spectrum at ILC has larger energy fractions in comparison with the WWA spectrum at CEPC. Second, the contribution of the single parton fragmentation (for instance the last Feynman diagram in Fig. \ref{feynman}) dominates the cross section in larger $p_t$ region at ILC, while the double parton fragmentation (for instance the first Feynman diagram in Fig. \ref{feynman}) dominates the cross section at relatively small $p_t$ region at CEPC.

\begin{figure}[!thbp]
    \centering
    \includegraphics[width=0.8\textwidth]{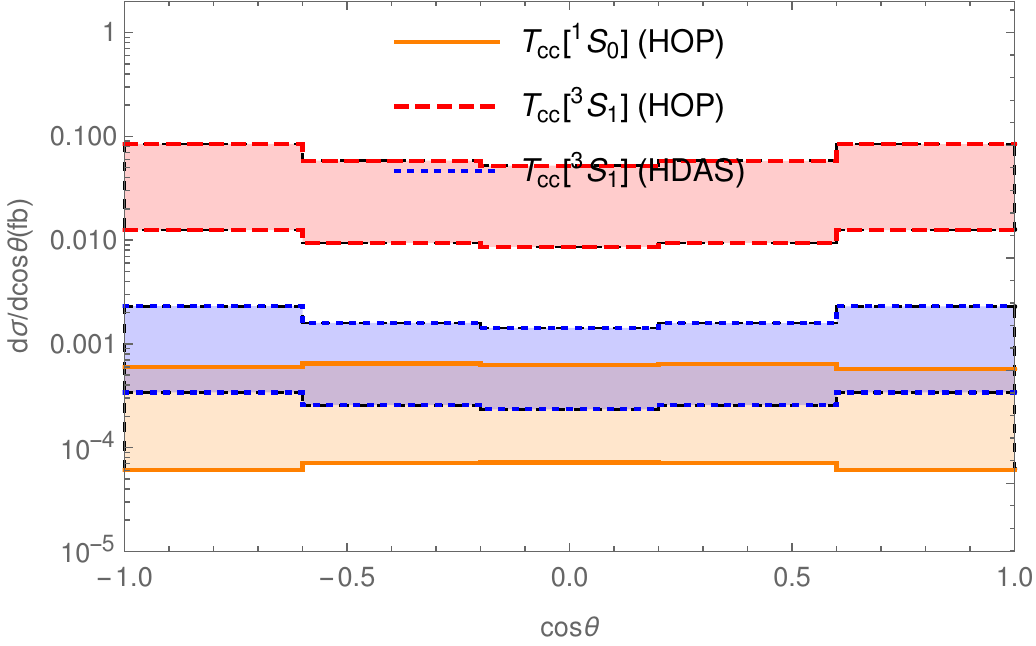}
    \caption{The differential angle distribution for the $T_{cc}$ production in two hadronization models with WWA photon spectrum at the SuperKEKB. Here $\sqrt{s} = 10.6\ {\rm GeV}$, the transverse momentum cut is $ 0.2\ {\rm GeV} \le p_{t} \le 4.0\ {\rm GeV}$, $m_{T_{cc}}=3.88$ GeV and the band is caused by the $m_c \in [1.5, 1.94]\ {\rm GeV}$.}
    \label{KEKBcos}
\end{figure}

\begin{figure}[!thbp]
    \centering
    \includegraphics[width=0.49\textwidth]{{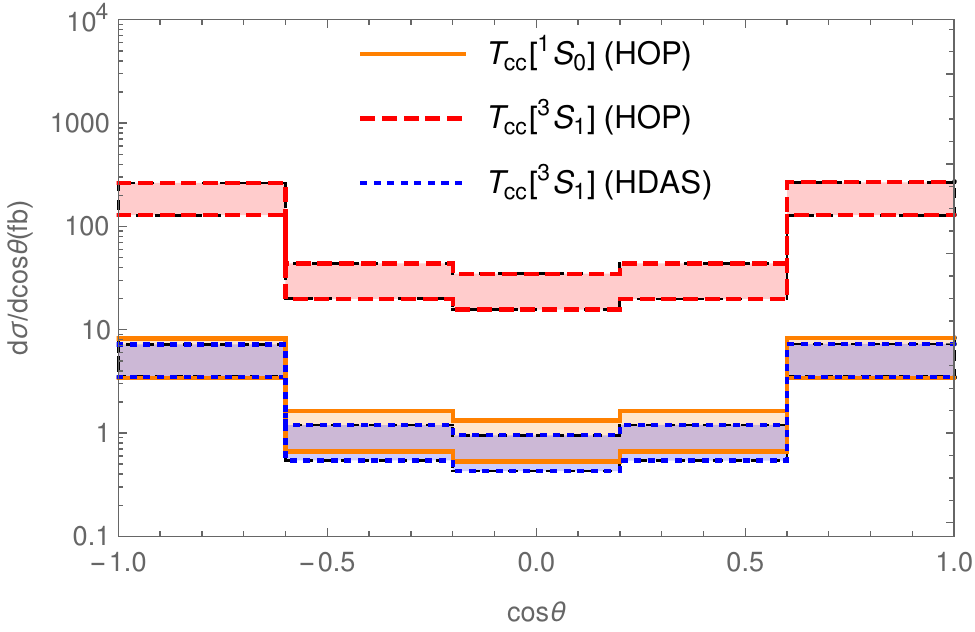}}
    \includegraphics[width=0.49\textwidth]{{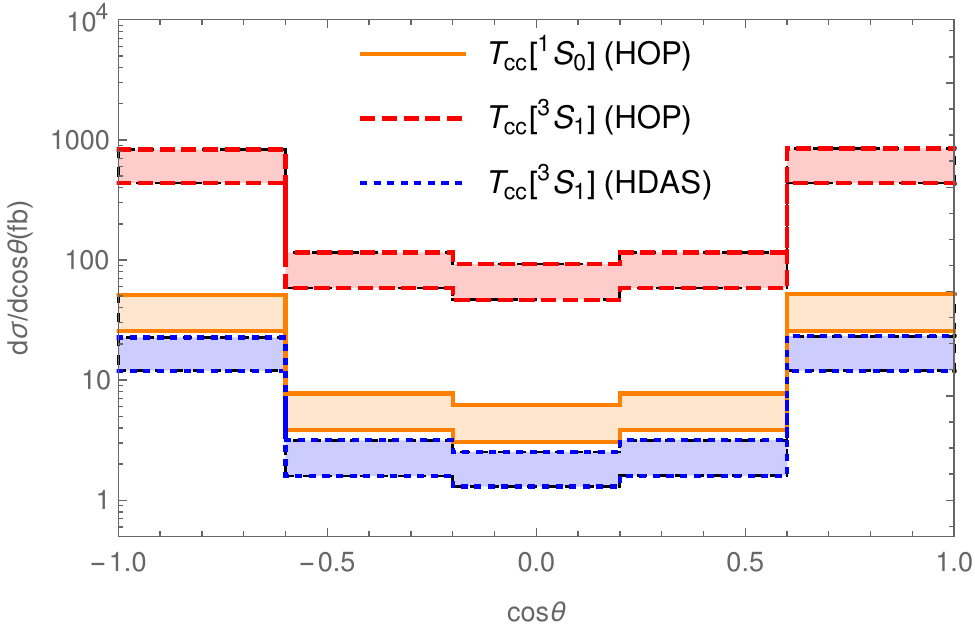}}
    \caption{The differential angle distribution for the $T_{cc}$ production in two hadronization models at the CEPC (left panel) and ILC (right panel). The transverse momentum cut is $ 1.0\ {\rm GeV} \le p_{t} \le 50.0\ {\rm GeV}$ at CEPC, while it is $ 1.0\ {\rm GeV} \le p_{t} \le 100.0\ {\rm GeV}$ at ILC. Here $\sqrt{s}=240\; {\rm GeV}$ for CEPC and $\sqrt{s}=500\; {\rm GeV}$ for ILC, $m_{T_{cc}}=3.88$ GeV and the band is caused by the $m_c \in [1.5, 1.94]\; {\rm GeV}$.}
    \label{CEPCcos}
\end{figure}

At last, we discuss the differential angle distributions d$\sigma$/dcos$\theta$ of $T_{cc}$, which are displayed in Figs. \ref{KEKBcos} and \ref{CEPCcos} for SuperKEKB and CEPC/ILC, respectively. 
Here $\theta$ is the angle between the momentum of initial photon and the momentum $q_1$ of the $T_{cc}$.
In Fig. \ref{KEKBcos} for SuperKEKB, it is shown that the differential angle distributions display a mild bulge for $[^1S_0]_6$ state but a mild valley for $[^3S_1]_{\bar{3}}$.
While in Fig. \ref{CEPCcos} for CEPC (left panel) and ILC (right panel), the differential angle distributions have an obvious valley when the $T_{cc}$ approaches the beam direction for both configurations.

\section{SUMMARY}
\label{sec:summary}
Within a phenomenological diquark fragmentation model, we study the inclusive production of tetraquark $T_{cc}$ via photon-photon fusion at $e^+e^-$ colliders. In the diquark model, both configurations of spin triplet color anti-triplet state $(cc)[^3S_1]_{\bar{3}}$ and spin singlet color sextuplet state $(cc)[^1S_0]_{6}$ for compact $T_{cc}$ tetraquark are considered.
At the SuperKEKB with $\sqrt{s} = 10.6\ {\rm GeV}$ and CEPC with $\sqrt{s} = 240\ {\rm GeV}$, the WWA photon spectrum for initial photons is adopted, while at ILC with $\sqrt{s} = 500\ {\rm GeV}$ the LBS photon spectra is taken into consideration.
In particular, two different hadronization schemes for the $(cc)[^3S_1]_{\bar{3}}$ diquark transformed into $T_{cc}$ are discussed. 
The total cross sections and the uncertainty caused by varying constituent charm quark mass of $(cc)$-diquark are displayed in Tabs. \ref{KEKB} and \ref{CEPC}. 
And the transverse momentum distributions of $T_{cc}$ and the differential angle distributions of $T_{cc}$ are presented in Figs. \ref{KEKBpt}-\ref{CEPCcos}.

We find that it is optimistic to observe the $T_{cc}$ signals as a compact tetraquark both at CEPC with CM energy $\sqrt{s} = 240$ GeV and typical luminosity of $10^{34}\ {\rm cm}^{-2}{\rm s}^{-1}$ and at ILC with $\sqrt{s} = 500$ GeV and the luminosity of $2 \times 10^{34}\ {\rm cm}^{-2}{\rm s}^{-1}$.
And the $(cc)[^3S_1]_{\bar{3}}$ diquark configuration dominates the $T_{cc}$ production under the same hadronization schemes.
It is found that the cross sections are sensitive to the constituent charm quark mass of $(cc)$-diquark, and also have strong dependence on the hadronization schemes.

\vspace{1.4cm} {\bf Acknowledgments}
This work is supported in part by National Natural Science Foundation of China under the grants 
No. 12235008, No. 12321005, No. 12275157, No. 12475083, No. 12475143, 
and by Natural Science Foundation of Shandong Province under grant No. ZR2023MA013.


\end{document}